\documentclass[prl,twocolumn,showpacs,superscriptaddress]{revtex4-1}
\usepackage{amsfonts}
\usepackage{epsfig}
\usepackage{bm}
\usepackage{amsmath}
\usepackage{multirow}
\usepackage{hhline}
\usepackage{graphicx}
\usepackage{bm}
\usepackage{dcolumn}
\usepackage{amssymb}
\usepackage[colorlinks,linkcolor=blue,citecolor=blue]{hyperref}
\usepackage{epsfig}
\begin{document}

\title{Electrostatic tailoring of magnetic interference in quantum point contact ballistic Josephson junctions}

\author{M. Amado}
\email{mario.amadomontero@sns.it}
\affiliation{NEST, Istituto Nanoscienze-CNR and Scuola Normale Superiore, I-56127 Pisa, Italy}
\author{A. Fornieri}
\affiliation{NEST, Istituto Nanoscienze-CNR and Scuola Normale Superiore, I-56127 Pisa, Italy}
\author{F. Carillo}
\affiliation{NEST, Istituto Nanoscienze-CNR and Scuola Normale Superiore, I-56127 Pisa, Italy}
\author{G. Biasiol}
\affiliation{CNR-IOM, Laboratorio TASC, Area Science Park, I-34149 Trieste, Italy}
\author{L. Sorba}
\affiliation{NEST, Istituto Nanoscienze-CNR and Scuola Normale Superiore, I-56127 Pisa, Italy}
\author{V. Pellegrini}
\affiliation{NEST, Istituto Nanoscienze-CNR and Scuola Normale Superiore, I-56127 Pisa, Italy}
\author{F. Giazotto}
\email{f.giazotto@sns.it}
\affiliation{NEST, Istituto Nanoscienze-CNR and Scuola Normale Superiore, I-56127 Pisa, Italy}
\begin{abstract}
The magneto-electrostatic tailoring of the supercurrent in quantum point contact ballistic Josephson junctions is demonstrated. An etched InAs-based heterostructure is laterally contacted to superconducting niobium leads and the existence of two etched side gates permits, in combination with the application of a perpendicular magnetic field, to modify continuously the magnetic interference pattern by depleting the weak link. For wider junctions the supercurrent presents a Fraunhofer-like interference pattern with periodicity $h/2e$ whereas by shrinking electrostatically the weak link, the periodicity evolves continuously to a monotonic decay. These devices represent novel tunable structures that might lead to the study of the elusive Majorana fermions.\end{abstract}

\pacs{73.21.Fg; % Quantum wells
      85.25.Cp;   % Josephson devices
      85.35.Be.   % Quantum well devices (quantum dots, quantum wires, etc.)
}
\maketitle

The interest into the electrostatic manipulation of the Josephson current in hybrid superconductor-normal metal-superconductor (SNS) nanostructures is attracting a great deal of attention in the last few years also thanks to the interest in topological superconductors,~\cite{Oreg-PRL,Beenakker-NJP} i.e., systems that might host the striking Majorana fermions.~\cite{Alicea,Beenakker-arXiv} To this end, proximized semiconductor-nanowires-based systems have been recently explored.~\cite{Mourik-Sci,Rokhinson-Nat,Deng-NanoLett,Das-Nat} Yet, ballistic two dimensional electron gas (2DEG) Josephson junctions (JJs) might be more suited for the investigation of this elusive particle.~\cite{Beenakker-NJP,Beenakker-arXiv}

The first evidence of the electrostatic tailoring of the Josephson coupling in proximized ballistic InAs-based nanostructures~\cite{Schapers-book} was reported by Takayanagi {\it et al.}~\cite{Takayanagi-PRL} In this work, the Josephson supercurrent passing through a superconducting quantum point contact (QPC) was modified by applying a negative voltage to the two splits gates, generating a depletion region in the nanoconstriction. This approach led to the reduction of the supercurrent and to the observation of plateaus of conductance. This behavior, predicted in~\cite{Houten} was confirmed experimentally in a subsequent work by Bauch {\it et al.}~\cite{Bauch-PRB} where the steps in conductance and the pinch-off of the QPC was seen in a sample similar to the one in Ref.~\onlinecite{Takayanagi-PRL}.

Heida {\it et al.}~\cite{Heida-PRB} performed the first characterization of the critical current ($I_J$) versus perpendicular-to-the-plane magnetic field ($B$) on 2DEG InAs-based SNS junctions showing a Fraunhofer-like interference pattern of the maximum of the supercurrent as a function of the external magnetic field. More recently, Harada {\it et al.}~\cite{Harada-PC} reported the first observation of $h/2e$ and $h/e$ periodicities in the interference pattern of physically different SNS junctions with different widths. They found that for wider junctions the periodicity of the $I_J-B$ behaviour obeys an ideal Fraunhofer-like pattern with $h/2e$ periodicity whereas for narrower SNS junctions the periodicity changes to a monotonic-like decay. These results have been lately confirmed either experimentally~\cite{Rohlfing-PRB} and theoretically~\cite{Cuevas-PRL}.
%%%%%%%%%%%%%%%%%%%%%%%%%%%%%
%%%%%%%%%%%%%%%%%%%%%%%%%%%%%
%%%%%%   F I G U R E  %%%%%%%
%%%%%%%%%%%%%%%%%%%%%%%%%%%%%
%%%%%%%%%%%%%%%%%%%%%%%%%%%%%
\begin{figure}[t!]
\centerline{\includegraphics[width=\columnwidth,clip=]{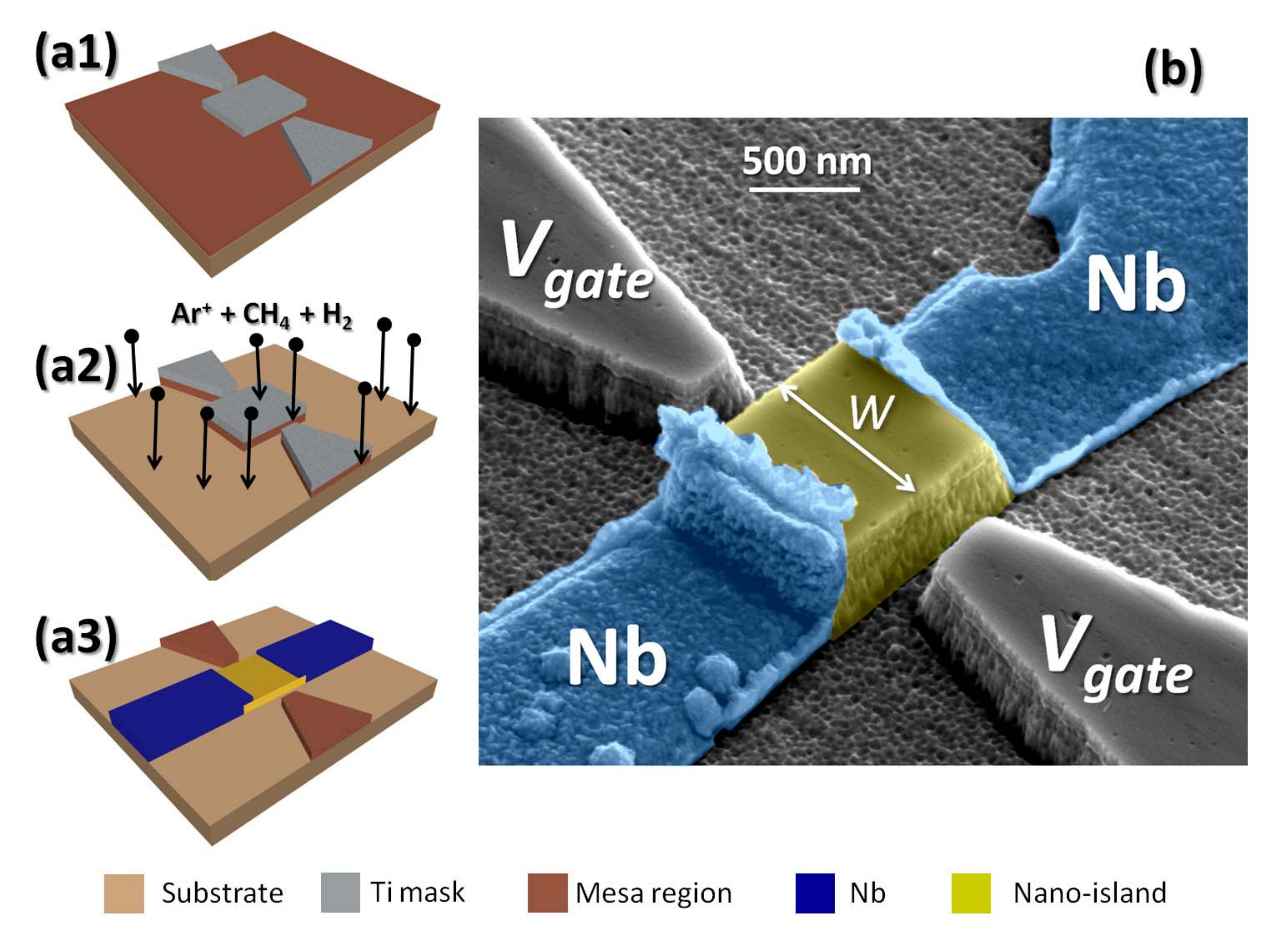}}
\caption{Sketch of the nanofabrication steps from the deposition of the Ti-mask (a1), reactive ion etching and definition of the mesa (a2), to the final device consisting of an InAs 2DEG QPC connected to Nb leads with two-etched lateral gates (a3). (b) Pseudo-color scanning electron micrograph of a typical superconducting QPC device. Nb contacts appear in blue whereas the InAs-region is in dark yellow.}\label{sample}
\end{figure}

Here we report the fabrication of InAs-based QPC ballistic JJs where the magnetic interference pattern of the supercurrent can be tailored at will by shrinking electrostratically the QPC. We demonstrate how the combination of the JJ and lateral gates allow us to modify continuously (i.e., within the same device) the interference period of the critical current from a Fraunhofer-like pattern to a monotonic decay one, thus exploring both the wide and narrow-junction limits. The nano-devices presented in this work appear as promising candidates for the generation of magneto-electrostatically finely-tunable ballistic JJs systems which could pave the way for the study of Majorana fermions (MFs).~\cite{Oreg-PRL,Beenakker-NJP,Alicea,Beenakker-arXiv} In sharp contrast with the widespread use of quantum wires (of InSb~\cite{Mourik-Sci,Rokhinson-Nat,Deng-NanoLett} and InAs~\cite{Das-Nat}) for the investigation of MFs where the systems are still far to be ballistic and the geometry is restricted to a simple wire, here we present an alternative approach based on the use of 2DEG InAs-based systems with a high {\it g}-factor. This approach permits the realization of micrometric-size ballistic junctions whose geometry can be tailored at will.

The heterostructure used to fabricate the JJs consists of an InAs quantum well (QW) grown by means of molecular beam epitaxy on the top of a (001) GaAs substrate with an intercalated series of $50$-nm-thick In$_{1-x}$Al$_x$As (with $x$ ranging from $x=0.75$ to $x=0.25$) essential to relax the strain between the substrate and the QW and to obtain low-defect-density and high mobility electron gases.~\cite{Capotondi-Crystal} The $4$-nm-thick InAs QW is finally inserted between two $5.5$-nm-thick In$_{0.75}$Ga$_{0.25}$As layers and In$_{0.75}$Al$_{0.25}$As barriers.~\cite{Capotondi-Thin} The sheet electron density turned out to be $n\simeq6.24\times10^{11}$ cm$^{-2}$ and the mobility in the dark $\mu\simeq1.6\times10^5$ cm$^2$/Vs, yielding a large elastic mean free path of $l_0\simeq2.3~\mu$m.

\begin{table}
%\begin{tabular}{c|c|c|c|c|c|c|c|c}
\begin{tabular}{c c c c c c c c c}
\multicolumn{9}{c}{} \\\hline\hline
\multirow{2}{*}{Device}&Width&$R_{N}$&$R_{sh}$&$I_{exc}$&$\mathcal{Z}_{exc}$&$\mathcal{Z}_{Sh}$&$\mathcal{T}_{exc}$&$\mathcal{T}_{Sh}$\\
 & nm & $\Omega$ & $\Omega$ & nA &  &  & $\%$ & $\%$\\
\hline
%Device & W (nm) & $R_{N} (\Omega)$ & $R_{sh} (\Omega)$ & $I_{exc}$ (na) & $\mathcal{Z}_{exc}$ & $\mathcal{Z}_{sh}$ & $\mathcal{T}_{exc}\%$ & $\mathcal{T}_{sh}\%$\\\hline
\multirow{1}{*}{A} & 800 & 714 & 256 & 377 & 0.96 & 0.94 & 52 & 52\\
\multirow{1}{*}{B} & 800 & 757 & 256 & 327 & 0.97 & 0.99 & 52 & 51\\
\multirow{1}{*}{C} & 600 & 961 & 341 & 292 & 0.95 & 0.95 & 52 & 52\\
\multirow{1}{*}{D} & 600 & 1150 & 341 & 242 & 0.96 & 1.09 & 52 & 46\\ \hline\hline
\end{tabular}
\caption{Physical parameters of the different devices presented in this work. $R_N$ is the junction normal-state resistance, $R_{Sh}=h\pi/2e^2W\sqrt{2\pi n}$ the Sharvin resistance (being $h$ the Planck's constant). Performing the calculation for the excess current ($I_{exc}$) as done in Ref.~\onlinecite{Deon-PRB} it is possible to obtain the scattering parameter $\mathcal{Z}_{exc}$ based on the Octavio-Tinkham-Blonder-Klapwijk model~\cite{Flensberg-PRB} and $\mathcal{Z}_{sh}$ from the relation $R_N=R_{Sh}(1+2\mathcal{Z}_{Sh}^2)$. The average in the normal state can be straightforwardly estimated as $\mathcal{T}_{exc,Sh}=(1+\mathcal{Z}_{exc,Sh}^2)^{-1}$.}
\label{samples}
\end{table}
%%%%%%%%%%%%%%%%%%%%%%%%%%%%%
%%%%%%%%%%%%%%%%%%%%%%%%%%%%%
%%%%%%   F I G U R E  %%%%%%%
%%%%%%%%%%%%%%%%%%%%%%%%%%%%%
%%%%%%%%%%%%%%%%%%%%%%%%%%%%%
\begin{figure}[t!]
\centerline{\includegraphics[width=\columnwidth,clip=]{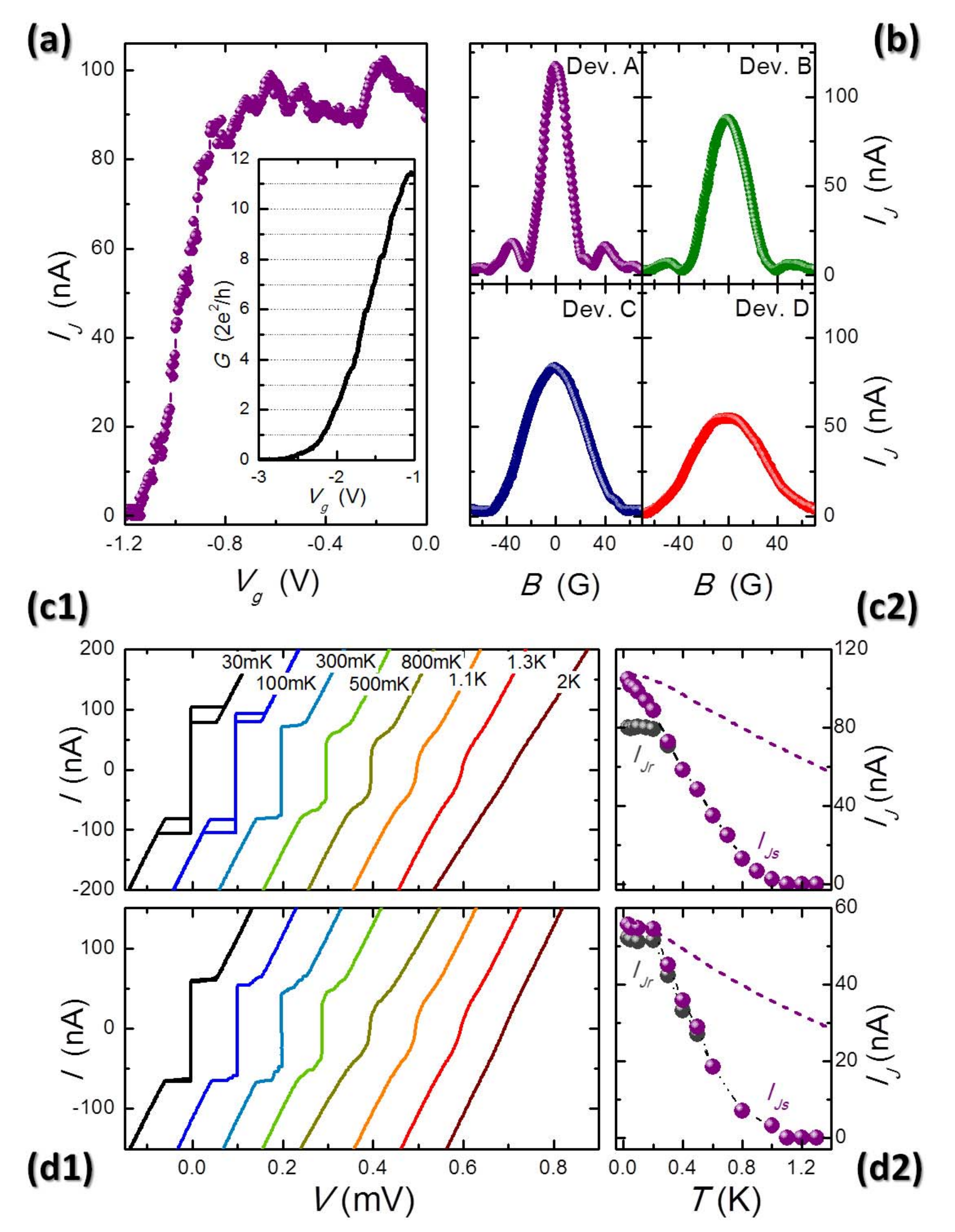}}
\caption{(a) Josephson current ($I_J$) vs gate voltage ($V_{g}$) with $B=0$ for device A. The Josephson coupling decreases by shrinking the QPC and vanishes for $V_{g}\leq-1.2\,$V. The normalized conductance as a function of $V_{g}$ measured at $B=125\,$G is displayed in the inset where the quantization of the conductance is hinted. (b) Magnetic interference patterns $I_J-B$ of all the four devices with $V_{g}=-0.5\,$V and $T\sim10\,$mK. Temperature evolution of a few selected SNS $I-V$ characteristics (at $B=0$ and $V_{g}=0$) in device A (c1) and D (d1). The curves have been horizontally shifted and the temperature spans from $30\,$mK to $2\,$K. Evolution of the switching ($I_{Js}$) and retrapping ($I_{Jr}$) supercurrents as a function of $T$ in device A (c2) and D (d2). Dashed lines show the  temperature dependence of the critical current calculated according to the model developed of Ref.~\onlinecite{Chrestin-PRB} with $\mathcal{Z}=1$ (c2) and $\mathcal{Z}=1.1$ (d2).}\label{figure2}
\end{figure}
The nanofabrication of the hybrid S-QPC-S JJs required a series of three mutually aligned steps of electron beam lithography (EBL) sketched in Fig.\,\ref{sample}\,[(a1)-(a3)].~\cite{Fornieri} Ni/AuGe/Ni/Au ohmics contacts were obtained in a single EBL operation followed by a subsequent aligned EBL step in order to define the semiconductor mesa into a QPC geometry with two lateral side-gates. A Ti mask was deposited by thermal evaporation [see Fig.\,\ref{sample}\,(a1)] and the heterostructure was then etched through reactive ion etching in Ar$^{+}$/H$_{2}$/CH$_4$ atmosphere~\cite{Carillo-PE,Fornieri} [see Fig.\,\ref{sample}\,(a2)]. The Ti mask was then removed by a $1$:$20$ HF:H$_2$O solution sample rinse. The resulting QPC mesa had a typical length of $L\simeq1~\mu$m, smaller than the elastic mean free path $l_0$. The third and last aligned step of lithography [see Fig.\,\ref{sample}\,(a3)] was fulfilled to laterally contact the 2DEG with two Nb electrodes, that were deposited by sputtering (with a previous dip into a $1$:$30$ HF:H$_2$O solution and a low-energy Ar$^{+}$ cleaning of the surface) with a growth rate of $\sim1.5\,$nm/s. A pseudo-color scanning electron micrograph of one typical device is shown in Fig.\,\ref{sample}\,(b). The superconducting gap of the Nb leads, of $\Delta_{Nb}\sim1.2$meV, yields a value of the superconducting coherence length $\xi_0=\hbar v_F/2\Delta\sim230\,$nm (the Fermi velocity $v_F\sim9.94\times10^{5}\,$m/s). Low temperature magneto-electric characterization was performed in a filtered dilution refrigerator down to $10\,$mK. 4-wire measurements were performed with the structure current-biased. The voltage drop across the junction was measured with a room-temperature differential preamplifier. Moreover, an external magnetic field was applied perpendicularly to the 2DEG plane to explore the Josephson current interference. The two side gates were used to deplete the 2DEG region of the QPC by biasing them negatively in voltage.

Table\,\ref{samples} shows the essential physical parameters of all four devices (A to D) studied in this work. The transmissivity ($\mathcal T$) has been indirectly estimated both from the excess current and the Sharvin resistance and presents a consistent value $\mathcal{T}\sim50\%$ in all the devices.

Figure\,\ref{figure2}\,(a) displays the evolution of the Josephson critical current ($I_J$) as a function of the gate voltage ($V_g$) measured without external magnetic field. The Josephson coupling decreases noticeably by shrinking the QPC and disappears for $V_{g}\leq-1.2\,$V. The averaged conductance ($G$) in device A, extracted from $20$ different $G-V_{g}$ characteristics, is shown in the inset. $G$ was measured in the regime of Josephson coupling suppression, at $B=125\,$G. Some conductance plateaus can be noted and the full suppression of $G$ is obtained when driving the QPC into the pinch-off regime at $V_{g}\leq-2.7\,$V. The other devices display similar characteristics. Figure\,\ref{figure2}\,(b) shows the evolution of $I_J$ vs. $B$ for all the four devices. Devices A and B, which have a width $W\sim800\,$nm, show a Fraunhofer-like pattern whereas the narrower junctions (device C and D having a $W\sim600\,$nm) exhibit a monotonic decay interference pattern. Figure.\,\ref{figure2} displays a selection of the current-voltage characteristics for devices A (c1) and D (d1) with the temperature spanning from $30\,$mK to $2\,$K.~\cite{cooldown}. The Josephson coupling survives up to $T\sim1.1\,$K for both junctions whereas traces of superconductivity persists at greater $T$. The temperature evolution for the switching ($I_{Js}$) and re-trapping current ($I_{Jr}$) is shown in Fig.\,\ref{figure2}\,(c1) (device A) and (d1) (device D). A noticeable hysteretic behavior is present in device A and B for temperatures below $300\,$mK whereas in C and D $I_{Js}\sim I_{Jr}$ for almost the whole range of $T$. We observed that quasiparticle heating in the 2DEG region when the junction switches into the dissipative regime~\cite{Courtois-PRL} manifests mainly in the wider junctions. Dashed lines represent the theoretical temperature evolution of $I_{Js}$ accordingly to the model of Chrestin {\it et al.}~\cite{Chrestin-PRB} calculated with the nominal values of $\Delta_{Nb}\sim1.2\,$meV, $n$, $W$ and $L$. The only fitting parameter is the value of the interface scattering parameter $\mathcal{Z}$~\cite{Flensberg-PRB} ($\mathcal{Z}=1$ for device A and $\mathcal{Z}=1.1$ for device D) yielding transparency of the interfaces of $\sim50\%$, thus, in very good agreement with the values extracted experimentally from $I_{exc}$ (Table\,\ref{samples}). At low temperature, the model recovers the experimental value for the maximum supercurrent in both junctions while the reduction predicted theoretically by increasing $T$ is much less pronounced than that observed experimentally.
%%%%%%%%%%%%%%%%%%%%%%%%%%%%%
%%%%%%%%%%%%%%%%%%%%%%%%%%%%%
%%%%%%   F I G U R E  %%%%%%%
%%%%%%%%%%%%%%%%%%%%%%%%%%%%%
%%%%%%%%%%%%%%%%%%%%%%%%%%%%%
\begin{figure}[t!]
\centerline{\includegraphics[width=\columnwidth,clip=]{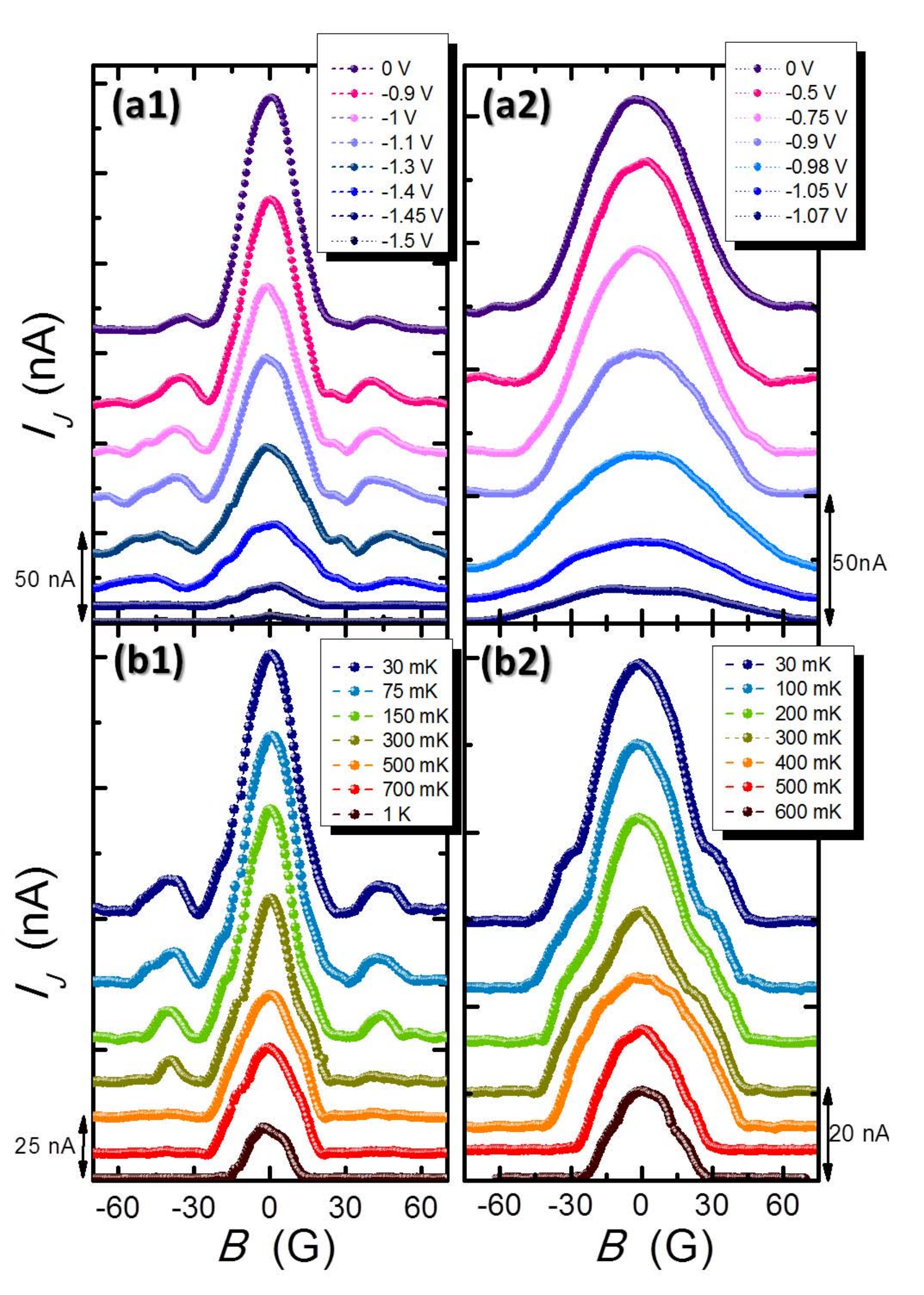}}
\caption{Magnetic interference pattern evolution of the $I_J-B$ characteristic for a few selected values of $V_{g}$ in devices A (a1) and C (a2). The curves have been vertically shifted and were recorded at $10\,$mK. Device A shows Fraunhofer-like interference pattern evolving to a the narrow-junction-like by depleting the weak link region. Device C shows the Fraunhofer-like interference pattern for $V_{g}\geq-0.5\,$V with up to two maxima visible and dramatically switches to the monotonic-like one when closing the QPC, enlightening the electrostatic tailoring of the magnetic interference in these ballistic S-QPC-S JJs. Evolution of the $I_J-B$ characteristic for a few selected values of $T$ in device A (b1) and D (b2). The curves have been vertically shifted and were recorded at $V_{g}=-0.75\,$V. Device A shows the Fraunhofer-like interference pattern with two maxima visible up to $T\leq300\,$mK remaining only the central peak when outrunning such temperature. Device D displays a monotonic-like interference pattern in the whole range of temperatures decaying faster than device A by increasing $T$.}\label{IvsB_Vg_T}
\end{figure}

Figure\,\ref{IvsB_Vg_T} shows the evolution of the $I_J-B$ characteristics measured at $10\,$mK for a few selected values of $V_g$ in devices A (a1) and C (a2). Device A exhibits the Fraunhofer-like interference pattern and by depleting the N-region (applying a negative voltage drop to the side gates) the pattern evolves to the one typically observed in narrow-junctions. Device C displays the Fraunhofer-like interference pattern for $V_{g}\geq-0.5\,$V with the second peak barely visible, and dramatically switches to a pattern characterized by a monotonic decay when shrinking the QPC further. From the position of the first zero in magnetic field of the $I_J-B$ pattern in Fig.\,\ref{IvsB_Vg_T}\,(a1) we can estimate the expected area of the proximized N-region from the expression $WLB=\Phi_0$, with $\Phi_0=h/2e=2.06\times10^{-15}\,$Wb the superconducting flux quantum. At $V_{g}=0\,$V $B$ turns out to be $\sim26\,$G thus $A\sim0.8\,\mu$m$^2$ that matches perfectly with the physical dimensions of device A and B. By depleting the QPC the position of the first minimum evolves to higher values of $B$. At $V_{g}=-1.45\,$V, the first minimum of the interference pattern manifests at $B\sim34\,$G yielding a value for $A\sim0.6\,\mu$m$^2$ which coincides with the area enclosed within N-region in the narrower junctions (device C and D). As we are depleting the N-region by electrostatic fields applied perpendicularly to the direction of the Josephson current, the evolution of the magnetic interference pattern is ascribed to the shrinkage of the effective width in the N-region.

%%%%%%%%%%%%%%%%%%%%%%%%%%%%%
%%%%%%%%%%%%%%%%%%%%%%%%%%%%%
%%%%%%   F I G U R E  %%%%%%%
%%%%%%%%%%%%%%%%%%%%%%%%%%%%%
%%%%%%%%%%%%%%%%%%%%%%%%%%%%%
\begin{figure}[t!]
\centerline{\includegraphics[width=0.9\columnwidth]{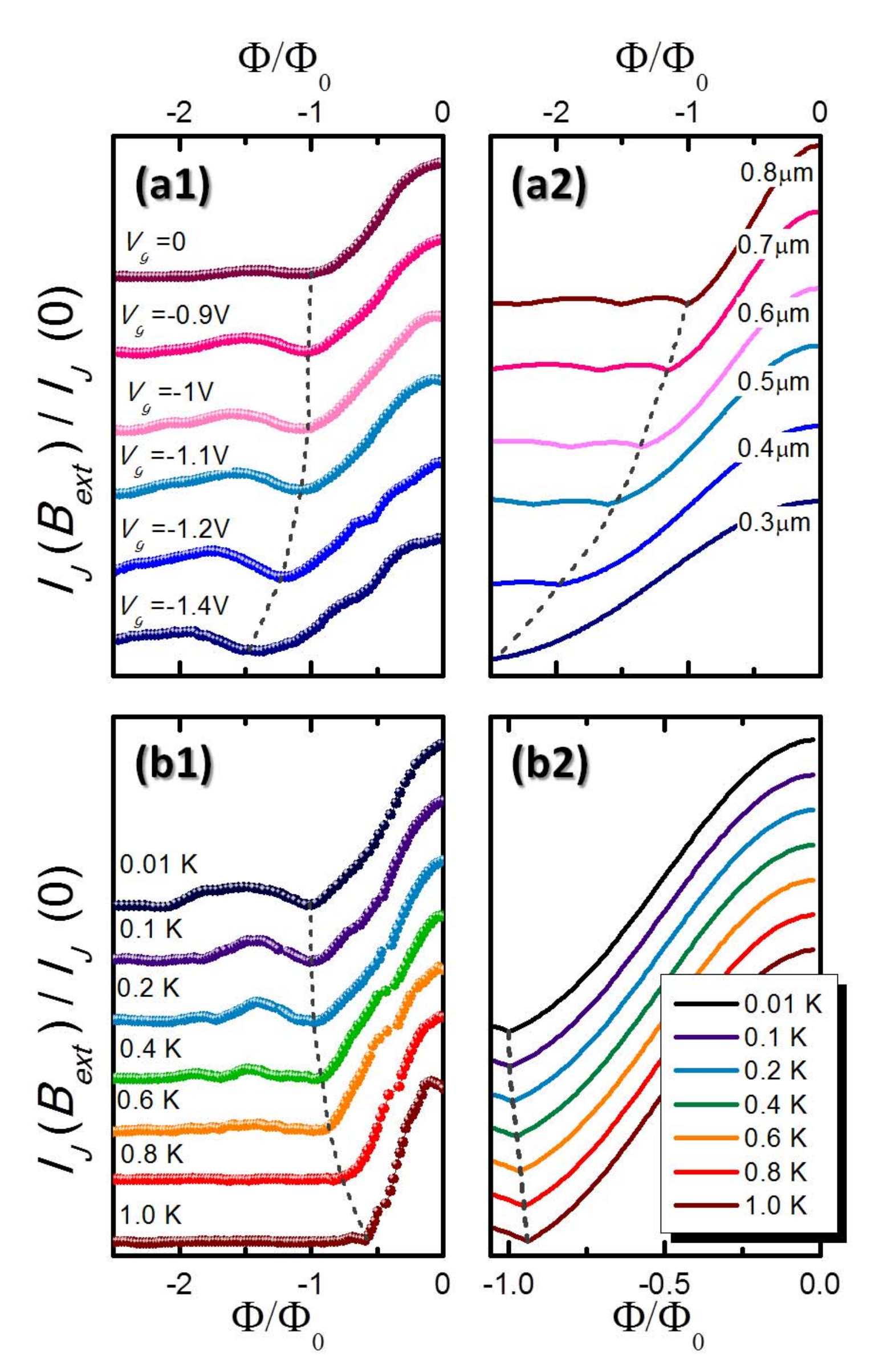}}
\caption{Evolution of the $I_J-B$ characteristic normalized to $I_J(0)$ for a few selected $V_g$ (a1) (measured at $10\,$mK) and $T$ (b1) (recorded at $V_{g}=-0.75\,$V). The {\it x}-axes have been normalized to the position of the first interference pattern minimum with $V_g=0\,$V (a1) and $T=10\,$mK (b1). Fit to Eq.\,\ref{Eq11Zagoskin} at diverse $W$ (a2) and $T$ (b2) [for the same temperatures that the shown in (b1) and $L=1\mu$m]. All the curves have been vertically shifted for clarity and the dashed-lines appear as guide for the eye stressing the position of the first minimum in the interference pattern. The {\it x}-axes have been normalized to the position of the first interference pattern minimum for $W=800\,$nm (a2) and  $T=10\,$mK (b2).}\label{I-vs-B-T-Zagoskin}
\end{figure}

To support our conclusions we now proceed to compare the experimental results with those obtained from a theoretical model introduced by Barzykin and Zagoskin~\cite{Barzykin}. This model predicts the evolution of the magnetic interference pattern as a function of the length-to-width ratio and temperature in ballistic mesoscopic SNS junctions. The expression for the Josephson current passing through the JJ is:
\begin{eqnarray}
I_J(\Phi)=\frac{2ev_{F}}{W\pi\lambda_{F}}\iint\limits_{-W/2}^{W/2} \frac{dy_1\,dy_2}{\left[1+\theta^2_{y_1-y_2}\right]^{3/2}}\times\nonumber\\
\sum\limits_{k=1}^\infty\frac{(-1)^{k+1}}{\xi_T\cos\theta_{y_1-y_2}}\frac{\sin k\left(\frac{\pi\Phi}{W\Phi_0}(y_1+y_2)+\phi\right)}{\sinh\frac{kL}{\xi_T\cos\theta_{y_1-y_2}}},
\label{Eq11Zagoskin}
\end{eqnarray}
where $\theta_{y_1-y_2}=\left(\frac{y_2-y_1}{L}\right)$ with $y_1$ ($y_2$) the position at the left (right) S-N interface, $\Phi=WLB$, $\lambda_f=31\,$nm and $\xi_T=\frac{\hbar v_F}{2\pi k_BT}$. Figure\,\ref{I-vs-B-T-Zagoskin}\,(a1) displays the experimental evolution of the $I_J-B$ characteristic normalized to the value at $B=0$ for a few selected values of $V_g$. In Fig.\,\ref{I-vs-B-T-Zagoskin}(a2) we show the evolution of the $I_J-B$ curves for different values of $W$ calculated from Eq.\,\ref{Eq11Zagoskin}. Gray dashed lines appear as a guide for the eye and show the evolution of the first interference pattern minimum. We note a good qualitative agreement between theory and experiment. The electrostatic shrinkage of the junction by the application of negative voltages in the side gates is equivalent to a physical reduction of the width of the junction, thus, a reduction of the number of conducting channels present in the system. By contrast, in Fig.\,\ref{I-vs-B-T-Zagoskin}\,[(b1)-(b2)] we display the normalized $I_J-B$ characteristics [experimental (b1) and theoretical (b2)] for a few selected values of $T$. Gray dashed lines appear again as a guide for the eye for the position of the first minimum of $I_J$. The theoretical curves obtained from Eq.\,\ref{Eq11Zagoskin} captures qualitatively the experimental evolution also as a function of temperature with the first minimum of the interference pattern displacing towards $\Phi_0$ when decreasing the temperature. The model, although rather idealized, is therefore an useful tool to grasp the physical picture at the basis of our junction's behavior.

In summary, we have reported the magneto-electrostatic tailoring of the supercurrent in ballistic nanofabricated JJ QPCs based on an InAs heterostructure laterally contacted to superconducting Nb leads. The depletion of the N-region by biasing negatively the two etched side gates allows us, in combination with the application of a perpendicular $B$, to tune the magnetic interference pattern continuously. We have demonstrated that for wider junctions the supercurrent displays a Fraunhofer-like interference pattern with periodicity $h/2e$ whereas by shrinking the QPC, the interference pattern evolves continuously up to a monotonic-like decay typical of narrower junctions as theoretically predicted.~\cite{Cuevas-PRL} The evolution of $I_J$ as a function of $W$ and $T$ in such samples has been qualitatively explained by a simple theoretical model that captures the underlying physics.~\cite{Barzykin}

We thank R. Aguado, D. Esteve, M. Goffman,  P. Joyez, H. Pothier and C. Urbina for fruitful discussions. Partial financial support from the FP7 program No. 228464 "MICROKELVIN", from the Italian Ministry of Defence through the PNRM project "TERASUPER" and for the POR CRO FSE 2007-2013 project "TERASQUID" is acknowledged.

%\newpage
%Density $n_s=6.24\times10^{15}\,$m$^{-2}$
%Mobility $\mu=16.6$ m$^2$/Vs
%Sheet resistance $R_{\square}\simeq60\Omega$. As our devices have a typical $L/W$ ratio $\sim1.25$ our 2DEG contributes to the total SNS resistance $\sim75\Omega$. The S-N resistance should be: $R_{S-N}=(R_{N}-75)/2\sim450\Omega$. As the QW has a thickness of $4\,$nm and the contact has a width of $1\,\mu$m, the S-N resistance is $\sim1\times10^{13}\,\Omega/$cm$^{-2}$
%Fermi energy = $64.2\,$meV
%$k_f=\sqrt{2\pi n_s}=1.98\times10^{8}\,$m$^{-1}$
%Effective mass = $m^*=0.023m_o$
%$\lambda_f=2\pi/k_f=31\,$nm
%Mean free path $L=v_f\tau=2.16\,\mu$m
%$\tau=\mu m^*/e=2.17\times10^{-12}\,$s
%$v_f=\hbar k_f/m^*=9.94\times10^{5}\,$m/s
%Superconducting coherence length
%$\xi_0=\hbar v_f/2\Delta\sim230\,$nm
%Induced coherence length, $\xi_N\sim\hbar^2\sqrt{2\pi n_s}/\Delta_{Nb}m^{*}\simeq 550\,$nm
%Thouless energy $\epsilon_{Th}\sim hv_F/L=4\,$meV.
%Length junctions $1\,\mu$m then $L\sim4\xi_0$ intermediate regime. Number Andreev bound states $L/\xi_0\sim4$
%If $\phi_0=h/2e=2.06\times10^{-15}\,$Wb, $\phi_0=AB$, $B=25\,$Gauss for width $W=800\,$nm, $B=32\,$Gauss for width $W=600\,$Gauss for the Fraunhofer-like pattern
%$R_N=R_{sh}(1+2\mathcal{Z}_{sh}^2)$

\begin{thebibliography}{99}
\bibitem{cooldown} The measurements were performed in different cool downs thus the electronic configuration of the 2DEG did not remain unchanged all along the realization of this work. Therefore, the maximum critical current found for a single device could slightly vary at a given temperature within the different figures.
\end{thebibliography}
\end{document}